\documentstyle{mn}
\title{Alignment of suprathermally rotating grains}
\author[A. Lazarian]
       {A. Lazarian\\Department of Astronomy, University of Texas,
Austin, TX 78712, USA\\
Harvard-Smithonian Center for Astrophysics, 60 Garden Street, Cambridge,
MA 02138, USA\\
DAMTP, Silver Street, University of Cambridge,
UK\thanks{Present address: Dept. of Astrophysical Sciences,
	 109 Peyton Hall,
	 Princeton University,
	 Princeton NJ 08544-1001, USA}}
\date{Received 25 April 1994
      Accepted }

\begin{document}
\maketitle

\begin{abstract}

It is shown, that mechanical alignment can be efficient for suprathermally
rotating grains, provided that they drift with supersonic velocities.
Such a drift should be widely spread due to both Alfv\'{e}nic waves and
ambipolar
diffusion. Moreover, if suprathermal rotation is caused by grain interaction
with a radiative flux, it is shown, that
mechanical alignment may
be present even in the absence of supersonic drift. This means that
the range of applicability of mechanical alignment is wider that it is
generally accepted and it can rival the paramagnetic one. We also study
the latter mechanism and reexamine the interplay between
poisoning of active sites and desorption of molecules blocking the access to
the active sites of H$_{2}$ formation in order to explain the observed
poor alignment of
small grains and good alignment of large grains. To have a more comprehensive
picture of alignment we briefly discuss the alignment by radiation fluxes
and caused by grain magnetic moments.
 \end{abstract}

\section{Introduction}

The polarimetry of the interstellar medium (hereafter ISM) testifies that  dust
grains are aligned.
The original explanation of this phenomenon in Davis
\& Greenstein \shortcite{dg} as being due to paramagnetic relaxation
was later criticized in Jones \& Spitzer \shortcite{js} as inadequate
 unless enhanced imaginary part of grain
magnetic susceptibility was assumed. At the same time, alignment involving
purely mechanical processes, pioneered in Gold (1951, 1952), was shown
to have its own problems (see Davis 1955, Purcell 1969, Purcell \&
Spitzer 1971). A new important contribution to the field was done by
Roberge \& Hanany (1990), who proposed an interesting idea, that grain
alignment can be caused by ambipolar diffusion. A comprehensive study of
alignment of thermally
rotating grains
through the ambipolar diffusion is given in
 Roberge, Messinger, \& Hanany (1995). Such an alignment is likely to be
important for molecular clouds. However it is possible to show, that
supersonic ambipolar diffusion cannot persist over extensive regions of the
ISM.

 A really profound step in elucidating
the dynamics of the ISM dust grains was done by Purcell (1975, 1979)
who introduced a concept of suprathermal rotation. However, it was
shown in Spitzer \& McGlynn \shortcite{sm} that this only marginally
improves the efficiency of paramagnetic alignment for high rates
of resurfacing. In fact,  specialists in the field believe
that `a relaxation mechanism orders of magnitude more efficient than
that given by normal paramagnetism' is required \cite{chg}.

This paper was preceded by our earlier papers devoted to the same issue.
First, in Lazarian (1994) (henceforth Paper I)
 we provided a quantitative description of mechanical
alignment of non-spherical grains. However, at that point it was unclear
whether grains can be aligned mechanically, if they rotate at suprathermal
velocities.
Therefore, only alignment of thermally rotating grains was
discussed. Then, in Lazarian (1995a) (henceforth Paper II)
 we addressed the problem of paramagnetic
alignment of suprathermally rotating grains and showed that diffusion of
oxygen over grain surface and existence of potential barrier for H$_{2}$
formation (Tielens \& Allamandola 1987) bring into being the grain critical
size $l_{cr}$
and the critical number of active sites $\nu_{cr}$. It was found out that
for grains with sizes less than $l_{cr}$ suprathermal
rotation due to H$_{2}$ formation is suppressed and that the poisoning
of active sites is accelerated when the number of active sites becomes greater
 than $\nu_{cr}$. However, only the case $\nu<\nu_{cr}$ was quantitatively
discussed and no processes of desorption were invoked; this is a disadvantage
of the latter study.

The aim of this paper is to determine roughly the relative importance of
processes, that can provide alignment of suprathermally rotating grains.
Therefore the
contribution is rather miscellaneous. First of all, in Sect.~2 we remind our
reader of the Purcell's concept of suprathermal rotation. Then,
we address in
Sect.~3 mechanical processes, which are usually believed to be incapable
of aligning suprathermally rotating grains. We show that this is not true and
discuss a process that can provide efficient alignment of such grains.
 In Sect.~4 we address the paramagnetic alignment and discuss poisoning
of grain active sites when
$\nu>\nu_{cr}$ and the processes of desorption of molecules blocking the access
to active sites.
 To provide a more comprehensive picture of alignment of
suprathermally rotating grains, we touch upon the alignment of helical grains,
grains subjected to
radiative flows, as well as alignment due to grain magnetic moments.

\section{Suprathermal rotation of grains}

Here we discuss the suprathermal rotation due to formation of  H$_{2}$
molecules on the ISM grains.
The number of molecules ejected per second from an individual site is
$\sim \gamma_{1} l^{2}n_{H}v_{1}\nu^{-1}$, where $\gamma_{1}$ is the
portion of H atoms absorbed by a grain of diameter $l$ while
$\nu$ is the number of
sites.  A recoil from a
molecule departing the grain is $m_{H2}v_{H2}=
(2 m_{H2} E_{H2})^{1/2}$, where $E_{H2}$ is
the kinetic energy of H$_{2}$ molecule ($0.2$~eV)
\cite{williams1}. Then the average squared $z$ component of torque is
\begin{equation}
\langle[M_{z}]^{2}\rangle\approx\frac{\gamma_{1}^{2}}{32}l^{6}n^{2}_{H}m_{H2}
v_{1}^{2}E_{H2}\nu^{-1}.
\label{7}
\end{equation}

Thus an individual grain spins up to high angular
velocities, limited only by friction forces. As a result, the attainable
value of angular velocity $\Omega$ is proportional to
$\langle[M_{z}]^{2}\rangle^{1/2}\kappa^{-1}$, where $\kappa$
is the constant of rotational friction $\sim I_{z}t_{d}^{-1}$. Note,
that $I_{z}$ is the $z$ component of the momentum of inertia and
$t_{d}$ is the rotational dumping time. The latter is
\begin{equation}
t_{d}\approx C t_{m}= C\frac{\varrho_{s}V}{S\Phi},
\label{e.4}
\end{equation}
where  $t_{m}$ is the time that takes a grain to collide with gaseous atoms
of the net mass equal to the mass of the grain, $V$ and $\varrho_{s}$ are
grain volume and density respectively; $S$ is the
grain  cross-section for a supersonic
flux, the coefficient $C=0.6$ is a precise result for a spherical grain
(Purcell \& Spitzer 1971) and $\Phi$ is the mass flux which is $v_{1}nm_{a}$
for supersonic $v_{1}$.
Therefore assuming $\gamma_{1}=0.2$ and $n_{H}=n$, one obtains
$\Omega\approx 2\cdot
10^{8}$~s$^{-1}$, for $\nu=10^{2}$.

Some grains are not subjected to the torques caused by H$_{2}$ formation.
For instance, it was shown in Paper II that those are
aromatic carbonaceous grains, as H$_{2}$ molecules are being
formed is states of
low excitation over their surface (Duley \& Williams 1993). Nevertheless, such
grains can rotate suprathermally, for instance, due to variations of the
accommodation
and/or photoelectric emission coefficients (see Purcell  1979). Further on,
we will show that unlike suprathermal rotation due to H$_{2}$ formation,
some of those processes may result in alignment on
their own.

Our estimates of $\Omega$ above are relevant to the long-lived spin-up
(Spitzer \& McGlynn 1979). For short-lived spin-up, namely, the correlation
time of Purcell rockets $t_{L}$ is much less than
$t_{d}$ our estimate of $\Omega$ must be multiplied
by $\sqrt{t_{L}/t_{d}}$. Speaking about Purcell alignment we will be mostly
concerned with the long-lived spin-up. Paramagnetic  alignment
corresponding to the short-lived spin-up does not differ much from
the Davis-Greenstein one corresponding to the enhanced temperature of
grain rotation $KT_{gas}$, where
\begin{equation}
K=\frac{I_{z}\langle[M_{z}]^{2}\rangle t_{d}t_{L}}{kT_{gas}}.
\label{3}
\end{equation}
It was pointed out by Bruce Draine (private communication), that such an
enhancement can be essential in molecular clouds where temperatures of
gas and grains are very close.

In fact,  suprathermal rotation can arise not only from
H$_{2}$ formation, but also by the variations of the
accomodation coefficient. We discuss these such processes in
Section~5, as we show, that such variations can cause mechanical
 alignment of
its own irrespectively from the action of the paramagnetic relaxation.

\section{Mechanical alignment}
\subsection{Supersonic motions}

To produce mechanical alignment grains should drift supersonically
in respect to gas.
In Paper I it is showed that Alfv\'{e}nic perturbations are efficient in
providing supersonic drift perpendicular to magnetic field lines. Two
main phenomena may be invoked:  Alfv\'{e}nic
waves and  ambipolar
diffusion.

It is generally accepted, that supersonic drift by ambipolar diffusion is
present within
sufficiently strong shocks (see fig.1 in Pilipp et al 1990). However, this
cannot be the cause of alignment for the majority of the ISM grains. Indeed,
it was pointed out by B.Draine, that the dissipation due to
ion-neutral streaming with velocity $v_{in}$ is approximately
\begin{equation}
G=0.5n n_{i} \langle \sigma_{T}v_{T}\rangle m_{a} v_{in}^{2}
\end{equation}
which for supersonic  $v_{in}=1$km s$^{-1}$ provides a dissipation rate per
H atom
$G/n \approx 1.7 \times 10^{-25}$erg s$^{-1}$, which is well in excess of
the energy that supernova is likely to eject into the turbulent motions.
In other words, the
ambipolar diffusion may be important for molecular clouds and localized regions
in within diffuse clouds, but for the large scale alignment other processes
must be invoked.

It was hinted in Paper I that large grains are inertial and therefore
should decouple from ionized gas in the course of
 high frequency Alfv\'{e}nic oscillations.
Although this process of alignment does not entail strong dissipation,
 it is possible to show that the limitations
imposed on  grain size by both damping of high frequency Alfv\'{e}nic waves
and grain charge are rather strict.

Indeed, if a grain without  electric
charge is placed  within partially ionized gas subjected to Alfv\'{e}nic
perturbations, e.g. with velocity $v=v_{0}\sin\omega_{A}t$, it
is easy to see that the
amplitude of grain drift scales as $v_{0}/\sqrt{1+\mu^{2}}$, where
$\mu \approx 0.4 (\rho_{n}/\rho_{g})(v_{0}/(l\omega_{A}))$ and $\rho_{n}$
and $\rho_{g}$ are, respectively, gas and grain densities.
 Therefore $\mu < 1$ provides the first constrain. In its turn,
Alfv\'{e}nic frequency cannot be arbitrary, as the waves of frequency higher
than that of collisional frequency for ions are critically damped (see
McKee et al 1994). The corresponding value of $\omega_{cr}$ provides the
minimal value of $\mu$. Another constrain stems from the grain charge, as if
$\omega_{A}$ is less than the grain Larmour frequency, the drift velocity
scales as $v_{0}(\omega_{A}/\omega_{L})$. For a given grain size
only oscillations within the range $\omega_{L}<\omega_{A}<\omega_{cr}$ are
efficient in the sense of providing supersonic drift. Therefore,
 even if the overall non-thermal line broadening of
emission lines is observed, it is not
clear a priori for a given grain, that this inequality is fulfilled
and that the Alfv\'{e}nic motions within aforementioned
range of frequencies
are supersonic. The estimates in Lazarian (1995b) show,
that for the ionization ratio $10^{-4}$
it is difficult to expect grains with radii less than $10^{-5}$ cm to
experience supersonic drift in the diffuse ISM, if grains are collisionally
charged (see Spitzer 1978, Draine \& Sutin 1987). It is gratifying, that
such a dichotomy is observed and grains of sizes less than
$10^{-5}$ cm are not aligned (Kim \& Martin 1994, 1995).

Radiation pressure can also cause grain drift. For this drift to be
supersonic
the following inequality $Q_{\it ext}P_{\it rad}>4P_{\it gas}$
should be satisfied (Purcell 1969), where $Q_{\it ext}$ is the ratio of the
optical to geometrical cross sections, while $P_{rad}$ and $P_{\it gas}$ are,
respectively, radiation and gaseous pressures.
For hydrogen with density 10~cm$^{-3}$ and temperature 100~K,
$P_{\it gas}\approx 10^{-13}$~dyn/cm$^{-2}$. Usually $Q_{\it ext}<1$ and this
means that it is difficult to obtain sufficient radiation pressure to account
for large-scale ISM polarization. Thus supersonic drift due to radiation
pressure
is confined to  regions in the vicinity of bright sources. As grains
carry charge, their motion under the radiation pressure is usually
constrained to following magnetic field lines (Spitzer 1978, p.~202).
It is shown in Habing {\it et al}. (1994) that grains should move
with supersonic relative velocities within the outflows around cool giants
(carbon-stars, Mira variables and OH/IR stars) due to
radiation pressure.  The velocities of metallic grains should be greater than
those of dielectric grains in such circumstances (Il'in 1994) and this may
have observational consequences in view of mechanical alignment.

	To summarize, supersonic drift is an essential feature of interstellar
grain
dynamics. In the vicinity of stars this
drift takes place mostly due to radiation pressure and happens along magnetic
field lines. Within molecular and atomic clouds at sufficient distances
from bright sources, this drift takes place due to Alfv\'{e}nic perturbations
and grains move mostly perpendicular to magnetic field lines.

\subsection{Cross section alignment}

Grain alignment can be quantitatively described by the following measure:
\begin{equation}
\sigma_{J}=\frac{3}{2}\left\langle\cos^2\beta-\frac{1}{3}\right\rangle,
\label{star}
\end{equation}
where $\beta$ is the angle between the angular momentum and the
alignment axis. The
latter can coincide with the direction of a corpuscular flux, if precession
on the time scale of alignment is negligible, or with the direction of
magnetic field, if the period of grain precession is much smaller than the
said time scale.
 For suprathermally
rotating grains the axis of major inertia is tightly coupled  to the
angular momentum (Purcell 1979). Therefore for oblate spheroidal
grains, $\sigma_{J}$ coincides with the Rayleigh reduction
factor $\sigma$ introduced in Greenberg (1968). For prolate spheroids it is
easy to show that
$\sigma=-0.5\sigma_{J}$.

One of the major questions that we need to answer here is whether
suprathermally rotating
grains can be aligned by a supersonic flow. A naive answer would be ''no``
as such rapidly rotating grains resemble gyroscopes and therefore
should not be sensitive to gaseous bombardment.
 However, this answer ignores the role of
crossovers. Further on we will show, that when
both random and regular torques are dominated by recoils caused by
H$_{2}$ formation,
 an efficient
alignment of non-spherical grains is possible.

We remind our reader, that during crossovers grain angular velocity becomes
close to zero and a substantial randomization of $\bf J$ is present.
A comprehensive theory of crossovers is given in Spitzer \& McGlynn (1979)
and further on we will use the results of this study.

 Let us assume,
for simplicity, that randomization is complete during a crossover,
which is apparently true for sufficiently small grains (see Paper II for
an explicit expression of the disorientation parameter). An incomplete
randomization only alters the time of alignment. For instance,
grain may need not one, but several crossovers to come to the state
uncorrelated with an initial one. This, however, does not change anything
for  our further treatment.

If $t_{L}$ is a characteristic correlation time for the existence of the
sites of H$_{2}$ formation, the mean time back to crossover is given by
\begin{equation}
t_{x}\approx 1.3(t_{d}+t_{L})
\label{5}
\end{equation}
(Purcell 1979). Usually, it is
assumed that the time of the existence of active sites is the time necessary
for accreting  a monolayer of refractory material (Spitzer \& McGlynn
1979). Another process that can determine $t_{L}$ is
 poisoning of the active sites (see Paper II and  Sect.~4 of this paper).
However, for our simplified treatment we do not distinguish between the
two processes limiting the life time of Purcell's rockets. Indeed, what is
important for us is that, for a chosen grain, $t_{L}$ is the time of accreting
of $N$ heavy atoms. The latter value depends on particular processes involved,
 but we will see that this is not critical for the mechanism below.

Consider at first a toy model, namely, assume
that the axis of suprathermally rotating oblate
grain can have only two positions, namely perpendicular ($\bot$) and parallel
($\parallel$) to the axis of a
gaseous flux. In this model, grain axis stays for the time $t_{x}$ in one
position and then undergoes the crossover and has equal chances to get
either  the same or the perpendicular alignment (disorientation is complete!).
In the
absence of the gaseous flux time scales $t_{x \parallel}$ and $t_{x \bot}$
are the same and, naturally, there is no preferential position. However, if
the gaseous supersonic
flux is present, $t_{d}$ given by Eq.~(\ref{e.4}) is inversely
proportional to gas-grain cross sections $S_{n \parallel}$
and $S_{n \bot}$ in the two positions. Moreover, it is natural to assume
that the
number of heavy atoms adsorbed by the grain is proportional to the overall
number of atoms striking it. Thus $t_{L}$ and $t_{x}$ are both
inversely proportional to the cross section. Assuming that the time of
the crossover is negligible as compared with the time of a spin-up, we
conclude that for an individual grain the time averaged probability
of finding the grain in a position $\bot$ or $\parallel$ is inversely
proportional, respectively, to $S_{n \bot}$ and $S_{n \parallel}$. To find the
constant of
proportionality one needs to recall,
that the probability of finding the
grain in either of two positions is unity. According to the
ergodic hypothesis this probability coincides with
the ensemble averaged one.

The above considerations can be generalized.
It is easy to see, that for a continues distribution of
axis positions
the probability of
finding grain axis within at a particular  angle is inversely proportional
to the cross section corresponding to this angle.

It was assumed above, that both $t_{L}$ and $t_{d}$ are controlled by the
same process, namely, by the
interaction of a gaseous flow with a grain. One may
imagine situations when $t_{L}$ and $t_{d}$ are controlled by {\it different}
physical processes. If one of the times $t_{L}$ or $t_{d}$ is much less than
the other,
\footnote{We should bear in mind that Eq.~(\ref{5}) is an approximate
one and it is checked in the range of $0.1<t_{L}/t_{d}<10$ (Purcell 1979).}
the longest of the two controls the alignment. For instance, for
$t_{L}\ll t_{d}$ we deal with alignment by friction, which may remind one
the the idea suggested in Salpeter \& Wickramasinghe (1969). However,
these processes are different, as we deal here with
 suprathermally rotating grains,
while  Salpeter \& Wickramasinghe discussed alignment of grains with
enhanced rotational temperature. It is possible to show, that the alignment
in the latter case is rather marginal. In the opposite limiting case, namely,
$t_{L}\gg t_{d}$ we show in Sect.~5 that alignment due to photodesorption
may be efficient. We do not dwell upon all these interesting possibilities
here,
as believe, that the alignment of suprathermally rotating grains under the
simultaneous action
of several processes deserves a separate study.

Consider the alignment of grains subjected to  Alfv\'{e}nic perturbations.
If, for the sake of simplicity, we approximate grains by thin discs
the cross section will vary as
\begin{equation}
S_{n}=\pi r^{2} |\sin\varphi \cos\psi|
\end{equation}
where $\varphi$ is the angle between the disc axis and magnetic field,
$\psi$ is an angle in the plane perpendicular to the magnetic field and $|..|$
denote that we take the absolute value of the trigonometric functions.
To account for the Larmour precession, we should perform averaging over
$\psi$. Therefore,
\begin{equation}
\langle \cos^{2}\varphi \rangle=C \int_{0}^{2\pi}\frac{d\psi}{|\cos\psi|}
\int_{0}^{\pi}\cos^{2}\varphi d\varphi
\end{equation}
where the normalization constant $C$ is
\begin{equation}
C^{-1}=\pi \int_{0}^{2\pi}\frac{d\psi}{|\cos\psi|}.
\end{equation}
In short, it is easy to see that $\langle \cos^{2}\varphi \rangle=0.5$ and
 the Rayleigh reduction factor
(Greenberg, 1968), is equal to $0.25$. Grain long axis tends to be
perpendicular to magnetic field, but the alignment is not perfect (compare
Paper I).

In another important case, when flakes stream along magnetic field lines,
similar computations provide $\sigma=-0.5$, which corresponds to the perfect
alignment with grain long axis along magnetic field.
 More efficient alignment for streaming along magnetic field
 as compared
with Alfv\'{e}nic perturbations
is a consequence of the fact, that in the former case
the Larmour precession
does not change the grain -- gas cross section.

It easy to see, that this type of
 alignment is efficient for flakes, as the cross section
difference averaged over the period of rotation is maximal for such grains.
Contrary to this,
the difference in cross sections
is not large for prolate grains. For instance, for needles
the
ratio of the maximal to minimal averaged
cross sections is just $\pi/2$ and the $\sigma$
 for the most favorable conditions (a flow parallel to the magnetic field
lines) is about $0.07$.
Therefore the alignment of prolate grains is marginal due to the mechanism.

It was implicitly assumed above that the angular momentum associated with
gaseous bombardment is not important
during a crossover. It is possible to show that, this is true unless grain
drift velocity is comparable with velocity of H$_{2}$ molecules and/or
accommodation coefficient is substantially different from unity and/or
atomic hydrogen is largely converted into molecular form.
A detailed discussion of the effects of gaseous bombardment
is given in Lazarian (1995c).

\section{Purcell alignment}

We will call paramagnetic alignment of suprathermally rotating grains the
Purcell alignment to distinguish it from the Davis-Greenstein mechanism
acting on thermally rotating grains. Various aspects of the Purcell alignment
 were addressed in Paper II.  Here we extend
 one aspect of the aforementioned
analysis, namely, the one dealing with poisoning of active sites.

Back in Paper II it was established that a critical number of
active sites exists. This number, $\nu_{cr}$, is the mean
number of sites with chemically adsorbed H atoms, that a
hydrogen atom arriving
to the grain surface can visit before reacting at any of these
sites. It is easy to see that, if the number of active sites is less than
$\nu_{cr}$, there cannot be on average more than one active site, which is
left empty since a recent H$_{2}$ formation. However, if the number of active
sites $\nu$ is greater than $\nu_{cr}$, the number of empty sites scales as
$\nu/\nu_{cr}$ for $\nu_{cr}\gg 1$. These empty sites are the primary targets
of
oxygen atoms
hopping over grain surface and it was observed in Paper II that
for $\nu>\nu_{cr}$ poisoning increases. Therefore only the case of
$\nu<\nu_{cr}$ was discussed.
Here we study the case of $\nu>\nu_{cr}$. Similarly to Paper II
we assume that oxygen is being immobilized on hydrogenation (Leitch-Devlin
\& Williams 1984, Williams, private communication).

If $\nu/\nu_{cr}$ is the expected number of empty active sites, the probability
of an oxygen atom to fill any of them as a result of an individual hop is
$\nu/(\nu_{cr}N_{ph})$, where $N_{ph}$ is the number of sites of physical
adsorption. Therefore the probability of an empty site not to be filled in one
hop of an oxygen atom is $(1-\nu/(\nu_{cr}N_{ph}))$; the same probability for
$m_{h}$ hops is
\begin{equation}
\left(1-\frac{\nu}{\nu_{cr}N_{ph}}\right)^{m_{h}} \approx \exp
\left(-\frac{\nu}{\nu_{cr}}
\frac{m_{h}}{N_{ph}}\right).
\label{ff}
\end{equation}

Therefore the characteristic time of poisoning of $\nu/2$ active sites is
\footnote{ We assume that not more than one oxygen
atom hop over grain surface at any
particular time. This may not be true for dense cores of molecular clouds,
where the concentration of atomic hydrogen is low.}
\begin{equation}
t_{p}=\nu t_{O}\frac{1}{1-\exp \left(-\frac{\nu}{\nu_{cr}}
\frac{m_{h}}{N_{ph}}\right)},
\end{equation}
where $t_{O}$ is the time of an oxygen atom arrival at grain surface.
For $\frac{\nu}{\nu_{cr}}\frac{m_{h}}{N_{ph}}\ll 1$ it is possible to expand
the exponent in Eq~(\ref{ff})
\begin{equation}
t_{p} \approx \nu_{cr} t_{O}
\frac{N_{ph}}{m_{h}}\left[1+\frac{1}{2}\frac{m_{h}}
{N_{ph}} \frac{\nu}{\nu_{cr}}-O\left(\frac{m_{h}^{2}}{N_{ph}^{2}}
\frac{\nu^{2}}
{\nu_{cr}^{2}}\right)\right].
\end{equation}

The number of hops $m_{h}$ is the ratio of the time required to hydrogenate
oxygen, which we denote $t_{ch}$ to the time of an individual hop
$t_{h}\approx 10^{-12}\exp(E_{h}/(kT_{s}))$ s, where $E_{h}$ is the energy of
potential barrier and $T_{s}$ is the grain temperature.
When the number of active sites exceeds $\nu_{cr}$ any hydrogen atom scans
only the fraction $\nu_{cr}/\nu$ of the entire surface before either reacting
with
another H atom at the active site or being trapped by an empty active site.
The probability, that O atom adsorbed by the grain is present over
this part of the surface is proportional to $\nu_{cr}/\nu$. Therefore the
time hydrogenation of oxygen is of the order of the $\nu/\nu_{cr}$ over the
timescale of hydrogen arrival. Therefore
\begin{equation}
t_{p}\approx \nu_{cr}\frac{\nu_{cr}}{\nu}\frac{t_{O}}{t_{H}}t_{h}N_{ph}
\left[1+\frac{1}{2}\frac{\nu^{2}t_{H}}{\nu_{cr}^{2}t_{h}N_{ph}}-
O\left(\frac{\nu^{4}
t_{H}^{2}}{\nu_{cr}^{4}t_{h}^{2}N_{ph}^{2}}\right)\right],
\label{12}
\end{equation}
where the ratio $t_{O}/t_{H}=(\gamma_{1}n_{H}v_{O})/(n_{O}v_{H})$; $n_{H}$ \&
$v_{H}$ and $n_{O}$ \& $v_{O}$ are the densities \& velocities of,
respectively,
hydrogen and oxygen. Note, that for thermal motions $v_{O}/v_{H}$ is equal
to $0.25$.

If the ratio  $\eta=\nu/N_{ph}$ stays constant for different grains,
it is evident from Eq.(\ref{12}) that $t_{p}$ at the first approximation is
$\approx 0.25\frac{\nu_{cr}^{2}}{\eta\gamma_{1}}t_{h}\frac{n_{H}}{n_{O}}$,
i.e. it is independent of the number of active sites for $\nu>\nu_{cr}$.
This is a new result, which was not foreseen in Paper II. This result means,
for instance, that the estimates of $t_{p}$
obtained there for
$\nu=\nu_{cr}$ are applicable to a wide range of grains with
$\nu>\nu_{cr}$.

Poisoning of active sites is essential for the Purcell alignment.
 A parameter that enters this
theory is the ratio of $t_{x}$ given by Eq.(\ref{5}) to the time scale of
paramagnetic relaxation $t_{mag}$ (see eq.~(58) in Purcell 1979, also see
Roberge et al 1993). To provide
sufficient alignment this ratio should be greater than unity (see fig.2 in
Purcell 1979). To obtain this for standard values of the ISM parameters
\footnote{It is argued in Paper II that these standard values may be misleading
for particular regions of diffuse clouds, but we avoid discussing
this issue here.} one needs to assume long-lived spin-up, i.e.
 $t_{L}\gg t_{d}$
(Spitzer \& McGlynn 1979) and therefore $t_{x}\sim t_{L}$. It is easy to
check that $t_{mag}$ scales as $l^{2}$, where $l$ is a grain size and therefore
on its own paramagnetic alignment favors small grains (see Johnson 1982).
For $\nu>\nu_{cr}$ in the first approximation we have obtained that $t_{p}$
scales as $l^{0}$ (see Eq.~(12)).

Note, that for $\nu<\nu_{cr}$ our arguments above are not applicable. The
corresponding study in Paper II showed, that in this case not more than
one active site is expected to be empty and the time of poisoning
is proportional to $\nu t_{O} N_{ph}/m_{h}$ for $m_{h}\ll N_{ph}$. The
number of hops for $\nu<\nu_{cr}$ is the ratio of time scale of the arrival of
a hydrogen atom $t_{H}$ and the time scale of the hop of the oxygen atom
$t_{h}$. Therefore both $t_{O}$ and $m_{h}$ scale as $l^{-2}$, while
both $N_{ph}$ and $\nu$ scale as $l^{2}$. As a result the time of poisoning
scales as $l^{4}$, which makes long-lived spin-up more probable for
large grains.

In general, the life time of Purcell's rockets $t_{L}$ is the minimal of
the following time scales: $t_{p}$ and the time scale of
accreting one monolayer of refractory
material. In diffuse clouds, as a rule, accreting of ice mantles is
suppressed (see Tanaka et al 1990). It is also believed, that hydrogenated
nitrogen and carbon within
diffuse clouds do not form mantles either. As the abundance of
heavier elements is negligible, it is natural to assume that in diffuse clouds
$t_{L}$ is controlled by poisoning of active sites.

Above we disregarded photodesorption. The characteristic time
for this process $t_{pd}$ scales as $l^{0}$ for smooth grains.
 The situation $t_{pd}\gg t_{p}$ corresponds to that studied in
Paper II. If $t_{pd}< t_{p}$, photodesorption removes molecules
blocking the access to active sites quicker that these sites are being
poisoned. Thus a long-lived spin-up is called into being.

As a result the following qualitative picture emerges. For small grains
poisoning dominates, i.e. $t_{pd}> t_{p}$, and therefore the spin-up is
short-lived. The time of poisoning for $\nu<\nu_{cr}$ grows with the
size as $l^{4}$ and for some size becomes equal to $t_{ph}$. Starting from
this
size we deal with long-lived spin-up; $t_{L}$ grows with the size until
$\nu=\nu_{cr}$. After the critical number of active sites is exceeded,
$t_{L}$ is stabilized at the attained level. In other words, marginal
alignment is expected for small grains and good alignment for large grains;
this corresponds to observations (see Kim \& Martin 1994, Kim \& Martin 1995).

Using both results obtained above and in Paper II we may attempt to
write equations for the dinamics of the number of active sites. Indeed,
while photodesprption, cosmic ray bombardment etc. create (or recover)
 active sites, poisoning removes them. The rate of active site creation
per unit area $A$ may be assumed constant, while the rate of
poisoning is different for $\nu<\nu_{cr}$ and $\nu>\nu_{cr}$.

If  $\nu<\nu_{cr}$ the results obtained in Paper II indicate, that the
rate of poisoning can be approximated by $CN_{ph}^{-1}$, where $C=
10^{12}\exp(-E_{h}/kT_{s})$ s$^{-1}$, while this rate is of the
order of $CN_{ph}^{-1}\nu^{2}\nu_{cr}^{-2}$ if  $\nu>\nu_{cr}$.
Therefore for  $\nu<\nu_{cr}$ the number of the active sites
changes as
\begin{equation}
\frac{{\rm d}\nu}{{\rm d} t}=A-C\frac{1}{N_{ph}},
\label{nu1}
\end{equation}
while for  $\nu>\nu_{cr}$ the following equation is valid:
\begin{equation}
\frac{{\rm d}\nu}{{\rm d} t}=A-C\frac{1}{N_{ph}}\frac{\nu^{2}}{\nu_{cr}^{2}}.
\label{nu2}
\end{equation}
Evidentely, Eq.~(\ref{nu1}) envisages a linear change of the number of
active sites. If the rate of desorption is greater than poisoning,
the number of active sites will increase linearly with time, unless all
the possible active sites are invoked over grain surface, provided, that
this number
is less than $\nu_{cr}$) or alternatively
 the number of active sites is reached
$\nu_{cr}$. In the opposite case when the rate of desorption is less than
the rate of poisoning the number of active sites decreases with time
untill all the active sites disappear. If  $\nu>\nu_{cr}$ the solution
of the Eq.~(\ref{nu2}) is as follows:
\begin{equation}
\nu=\nu_{cr}\sqrt{\frac{AN_{ph}}{C}}\frac{1+B\exp(-\frac{2}{\nu_{cr}}
\sqrt{\frac{AC}{N_{ph}}}t)}{1-B\exp(-\frac{2}{\nu_{cr}},
\sqrt{\frac{AC}{N_{ph}}}t)}.
\label{nu3}
\end{equation}
where B is related to the number of active sites $\nu_{in}$ at $t=0$ in the
following way:
\begin{equation}
B=\frac{\nu_{in}-\sqrt{\frac{A}{C}}}{\nu_{in}+\sqrt{\frac{A}{C}}}.
\end{equation}
Eq.~(\ref{nu3}) testifies, that the number of active sites
tend to stabilize at the level $\nu_{cr}\sqrt{AN_{ph}/C}$.
Therefore for $A$ a bit larger than $C/N_{ph}$, the number of
active sites is expected to be of the order $\nu_{cr}$ in correspondence
with a qualitative conclusion reached in Paper II.

Unfortunately, in our ignorance of many parameters involved, it is extremely
difficult to quantify this otherwise luring picture. First of all, we have
rather uncertain knowledge of the density of active sites as well as of other
parameters of grain surface.
Then, photodesorption presents another problem, as the rates obtained in some
laboratory
experiments (see Bourdon, Prince \& Duley 1982) are very low.
We are not sure either whether the photodesorption is driven by UV quanta
only or also by 3 $\mu$m radiation as it is claimed in Williams et al (1992).

In view of this ambiguities, we need to treat the picture above as a
conjecture only. We believe, that further research in the field will test
this conjecture.

An interesting feature of Eq~(\ref{12}) that it shows that $t_{p}$
is proportional to the number of sites of physical adsorption. Therefore
fractal grains with large surface area should correspond to larger
$N_{ph}$. The alignment of fractal grains was studied in Lazarian (1995d),
where the excess of the physical area for such grains over that for smooth
grains was invoked in order to improve the efficiency of alignment.
This study
indicated a possibility of a substantial improvement of alignment, provided
that molecules blocking the access to active sites had sufficient mobility.
 This assumption seems to be a shortcoming of the
latter study. However, it is possible to show that the
dependences for alignment measure on the fractal dimension
obtained in Lazarian (1995d) are valid
without this assumption just as the result of the dependence of $t_{p}$
on $N_{ph}$. Another worry in Lazarian (1995d) was, that
for fractal grains the majority of H$_{2}$ formation events
 may take place within
narrow pores, and this may suppress suprathermal rotation.
Our arguments above show, that the sites within grain pores
are likely to be poisoned and therefore H$_{2}$ formation should take place
mainly over open surfaces of grains, which are kept clean e.g.
due to photodesorption. At the same time, surfaces within pores provide oxygen
with more space for its random walk hopping.

To summarize, we have shown, that the Purcell (1979) mechanism analogously
to the Mathis (1986) one favors large grains.
 To tell these two mechanisms it is advantageous to
study dependences of degree of alignment on grain temperatures.
Our discussion above indicates, that $t_{p}$ is proportional to $t_{h}$,
which varies exponentially with temperature. Therefore the Purcell alignment,
unlike the Mathis one
should be very sensitive to changes of grain temperature.

Above we assumed that initially the concentration of active sites over
grain surface is high. Although this corresponds to the modern picture of
grain chemistry (see Tielens \& Allamandola 1987, Buch \& Zhang 1991),
an alternative approach to the problem is also possible. Indeed, if
the surface density of
active sites may be less than $10^{-5}$ cm$^{2}$ and than only grains
larger than $10^{-5}$ cm  are likely to have at least one active site
(Lazarian 1994b).\footnote{I suggested this possibility, but did not
treat it seriously untill it got support from John Mathis.}
This  idea requires further study, but here we would like
to point out, that the case of high and low density of active sites can
be distinguished by temperature dependence. It is easy to see, that
if the density of active sites is low, while nascent H$_{2}$0 molecules
are ejected on formation, the temperature dependence of the
Purcell alignment is suppressed.

\section{Other possibilities of alignment}

{\bf Alignment of helical grains.}
Grains studied above were symmetric.
 However, real grains may have helicity. In fact, grains
in Purcell (1975, 1979), that rotate suprathermally due to variations of
the accommodation coefficient
or photoelectric emission coefficient are helical. What was omitted in the
above study is that the two aforementioned effects can produce alignment even
without paramagnetic relaxation. For instance, radiation is, as a rule,
anisotropic and this may influence the rotation caused by photoelectric
emission.

There is a substantial difference when a helical grain is subjected to
a isotropic bombardment of atoms or photons and when it is subjected to a flux.
In the former case, there is no difference between H$_{2}$ formation
and other causes of suprathermal rotation; with high degree of accuracy
it is possible to assume, that the regular torque acts only along the axis of
major inertia (see Spitzer \& McGlynn 1979). If, however, a helical grain is
subjected to a radiative or
corpuscular flux, the component of torque perpendicular to the axis of
major inertia can cause
 precession, which may significantly alter the alignment.

An interesting example of
helical grains was discussed in Dolginov \& Mytrophanov (1976). There,
grains subjected to regular torques due to the difference in
scattering of left- and right- circular polarized photons were considered.
  It was  found, that if grains, either due to their
chemical composition or due to their shape, have different cross-sections for
left- and right-hand polarized quanta (see Dolginov \& Silantev 1976)
their scattering of unpolarized light results in the
spin-up. The efficiency of this process is maximal for
wavelengths $\sim l$. For twisted prolate grains,  increments of grain
angular momentum  can be shown \cite{dm} to be of the order of
$0.25\Phi_{p}\hbar l^{2}(n_{r}-1)^{4}$, where $\Phi_{p}$ is the
flux of photons and $n_{r}$ is the refraction coefficient ($\sim
1.3$). Then the characteristic angular velocity is
\begin{equation}
\Omega\approx\frac{1}{8}\frac{\Phi_{p}(n_{r}-1)^{4}\hbar}{nm_{a}v_{a}l^{2}},
\end{equation}
which is of the order $10^{7}$~s$^{-1}$ for the typical concentration
of photons in the ISM $\sim 3\cdot 10^{9}$~cm$^{-2}$~s$^{-1}$ and
may become much higher in the vicinity of bright sources.
For optimal shape suggested in Dolginov \& Mytrophanov (1976)
the scattering efficiency is sufficient to account for the alignment over
vast regions, but one
may expect the grains to have relatively small deviations towards the optimal
form. Such grains are believed to be a natural product of evolution
(Mathis 1990).
However, numerical studies of the interaction of irregular grains with
radiation by Bruce Draine (private communication) showed, that grains
that do not resemble helicies can exibit high efficiency of spin-up
due to scattering of radiation. These studies can shed the light on
the problem, whether grain spin-up arising from light scattering
 is an exeption or a rule. Note, that such a spin-up is expected to
be long-lived one. Indeed, it is not the surface, but grain volume,
that should be altered substantially to cause a crossover event.
Such a change is expected to take place over timescale much longer
as compared with the time of gaseous damping, which means that
the spin-up should be long-lived. For such a spin-up the alignment is
nearly perfect just due to  paramagnetic relaxation.

However, helical grains can be aligned mechanically on the timescale
much shorter than the one relevant to the paramagnetic alignment.
Mechanical alignment of helical grains is a complex problem and we
are going to
subject this issue to scrutiny in our next paper. Here we
just refer to the study in Dolginov \& Mytrophanov (1976) where it was shown,
that grains asymptotically tend to align perfectly with their helicity
axes along the direction of magnetic field, if the time of precession is
much less than that of alignment, and along the flux direction, if the opposite
is true. In their paper Dolginov \& Mytrophanov considered a tilted oblate
grain which helicity axis coincided with the major inertia axis and a tilted
prolate grain where the two axis were perpendicular. The conclusion reached
in Dolginov \& Mytrophanov (1976) was that the two species subjected to a flux
should be aligned orthogonally. However, the above study omits the influence
of internal relaxation. Due to this effect a prolate grain initially rotating
about its axis of minimal inertia
 will in a short period of time turn to rotate about the axis of major inertia.
In other words, long axis of prolate and oblate grains should be aligned in
the same direction.

To obtain quantitative results applicable to the ISM and circumstellar regions
one needs to estimate the relative importance of this particular type of
suprathermal rotation. Estimates in Purcell (1979) show that, as a rule,
angular velocity of rotation due to H$_{2}$ formation should dominate
for the ISM. Therefore, it seems unlikely that alignment due to
grain helicity is more important in diffuse
clouds in comparison with the one discussed in Sect.~3.
However, the alignment of helical grains may be essential in the vicinity of
bright sources.\\

{\bf Alignment due to radiation fluxes.}
Above we discussed both mechanical  and
paramagnetic alignment of suprathermally rotating grains.  One may
wonder whether alignment can be caused by absorption of quanta. Indeed,
due to absorption of an individual quantum a grain gains $\hbar$ angular
momentum. This physical process is invoked in Harwit (1970).
However, it was shown later in Purcell \& Spitzer (1971)
that if a grain is being subjected to a radiation flux, the drift produced
by the radiation pressure entails Gold alignment, which is far more efficient
than the Harwit one. It is not difficult to show, that the ratio of squared
increments of angular momentum for the Harwit process $\langle( \triangle
J_{H}^{2})\rangle$ to the one for the Gold process
$\langle( \triangle J_{G}^{2})\rangle$
is negligible even for rather special conditions of alignment
discussed in Aitken et al (1985). Note, that in the aforementioned
study the Gold alignment was neglected and therefore a conclusion, that
Harwit mechanism is efficient was reached. We claim, that this inference can
be true only, if the Gold alignment is suppressed. The latter happens, for
instance, if charged grains are trapped by close loops of magnetic field
and cannot be accelerated under radiation pressure. In any case, such
an inefficient process as a Harwit one, is unlikely to influence the
dynamics of suprathermally rotating grains.

This, however, does not mean,
that radiation cannot influence alignment of suprathermally rotating
 grains in a way other than
through differential scattering of circular polarized quanta. For instance,
we may suggest a mechanism based on photodesorption. Indeed, if suprathermal
rotation is due to H$_{2}$ formation, we may refer to the toy model discussed
in Sect.~3, but use a radiation flux instead of gaseous one. If this radiation
flux
desorbs molecules blocking the access to active sites, the life time of
Purcell's rockets and therefore $t_{L}$ will be different for $\bot$ and
$\parallel$
orientation of our grain. It is easy to see, that this alignment is most
efficient if $t_{L}\gg t_{d}$. However, in some cases, e.g. for hot and
small grains discussed in Purcell \&
Spitzer (1971), the radiation dumping may dominate the gaseous one. In this
case  the grain temperature should depend on the
grain orientation and $t_{L}\gg t_{d}$ is not required for efficient alignment.
 However, a
detailed study of these processes is beyond the scope of our present
paper.\\

{\bf Alignment due to grain magnetic moments}.
To make our discussion more complete, we need to mention the alignment due
to grain magnetic moments. One of the causes of these moments can be
charge (Martin 1971, Draine \& Salpeter 1979, Draine \& Sutin 1987)
 over rotating grains (Rowland effect). The torques that act on a rotating
charged grain were studied in Davis \& Greenstein (1951), along with their
famous prediction of paramagnetic alignment. However, the influence of the
ambient gas
was omitted in their analysis, and thus the conclusion that the
`distribution must be completely independent of the field' was made.
This was quoted in other sources  and therefore may still cause confusion.

Consider an isolated rotating charged grain in magnetic field. Its rotation
creates a magnetic moment  ${\bf \cal M}=
\Sigma_{i}{\rm e}(2c)^{-1}a_{i}^{2}{\bf J}({\it I})^{-1}$,
where $a_{i}$ is the distance from an elementary charge over the grain
surface to the axis of rotation. This causes grain precession
in the magnetic field  according to the Larmour equation
\begin{equation}
\frac{{\rm d}{\bf J}}{{\rm d}{\it t}}=
{\bf \cal M}\times{\bf B}=
\omega_{g} {\bf J}\times\frac{{\bf B}}{|
{\bf B}|}
\label{56}
\end{equation}
where $\omega_{g}=\sum \frac{{\rm e}a_{i}^{2}B}{3cI}$, which is quite close to
the grain Larmour frequency
$\omega_{L}$. In the course of this precession the angle between
${\bf \cal M}$ and ${\bf B}$ does not change,
i.e. the precession by
itself cannot change the energy of the system. The calculations
in Davis \& Greenstein (1951) reflect this fact. However,
 the energy changes due to the interaction of grains with surrounding
gas. Indeed,
the impact in the direction of the Larmour precession arising from
grain interaction with an atom results in
the torque, that  cause the precession around the axis which is
perpendicular to the magnetic field. Eq.~(\ref{56}) shows
that the angle between ${\bf \cal M}$ and
${\bf B}$  increases. If
the impact is directed against the direction the Larmour precession,
this angle decreases.\footnote{Note, that in reality
we assume a uniform distributions
of atomic impacts and subdivide the impacts as having component either
in the direction of the precession or opposite to it.}
 The collisions
of the first type disalign ${\bf \cal M}$ and
${\bf B}$, while the collisions of the second type align the two
vectors.  The alignment effect  prevails as the impacts against
precession are stronger than those in the direction of precession.

In our arguments above we used the fact, that the Larmour precession
influences the direction, but not the magnitude of the vector $\bf J$.
Therefore the magnitude of ${\bf \cal M}$ (i.e. $|{\bf \cal M}|$)
is not altered by the Larmour precession. In fact, in terms of
grain interaction with magnetic field, the grain behaves as a magnetic dipole
and the alignment due to the Rowland effect is similar to the alignment of
magnetic dipoles. Such an alignment is not complete due to random torques
acting upon grains.\footnote{Thermal fluctuations within grain material
can randomize alignment as well, but the influence of such fluctuations
for suprathermally rotating grains is negligible.} These random torques
determine the effective temperature $T$ and this temperature influences the
equilibrium distribution of ${\bf \cal M}$. If the suprathermal rotation
is caused by the variations of the accommodation coefficient
(see Purcell 1979),
 this temperature will be the mean of the grain and gas temperatures,
provided that gas-grain collisions are inelastic. In the case of the
suprathermal rotation caused by H$_{2}$ formation, $T$ will be of the
order of $E_{H2}/k$, provided that every hydrogen atom that hits
 grain surface leaves it as a part of H$_{2}$ molecule.

 Indeed, if the spin-up is very long, then grains should behave
as magnetic moments whose motion is disturbed by stochastic torques due to
H$_{2}$ formation.\footnote{Note, that we are speaking about averaged
$J^{2}$ components of the torque.} These magnetic
moments are peculiar in a sense, that the randomization of their precession
requires time much greater than usual damping time. However, if the life time
of Purcell's rockets exceeds this time, the treatment in mu228rv2 should be
applicable.

In the extreme case of the short spin-up, when all torques are
essentially stochastic one may introduce the averaged magnetic moment and
study
precession of this moment in the magnetic field. Our arguments
should be applicable and the damping of the precession should occur over
the timescale of the order of rotational damping time.

 The equilibrium distribution of grain axes of major inertia
in magnetic field $B$ is proportional to $\exp(- \mbox{$\cal E$}/(kT))$,
where ${\cal E}=-{\bf \cal M}{\bf B}$.
It is easy to see that
\begin{equation}
\frac{\mbox{$\cal E$}}{kT}\approx
\frac{\rm e}{2c}l^{2}\frac{\omega_{T}B}{kT}
\sim\frac{\omega_{L}}{\omega_{T}}
\label{54}
\end{equation}
for a thermally rotating grain  and
$\mbox{$\cal E$}(kT)^{-1}\approx
\omega_{L}\omega_{r}\omega_{T}^{-2}$ for a suprathermally rotating grain,
$\omega_{T}$ is the frequency of
rotation corresponding to the temperature $T$,
$\omega_{r}$ is the  frequency of the suprathermal rotation and $\omega_{L}$
is the Larmour frequency.

In fact, grain alignment due to the Rowland effect is in no way
different from the alignment of ferromagnetic grains with high permanent
magnetization discussed in Spitzer \& Tukey (1951). The difference
is of quantitative nature; ferromagnetic grains have much larger magnetic
moments and therefore much more susceptible for such an alignment. However,
even with ferromagnetic grains the alignment was shown to be not efficient for
diffuse clouds \footnote{ It is possible to show that Spitzer \& Tukey
alignment may not be negligible in dark clouds, where both grain and gas
temperatures are low, while magnetic fields $> 10^{-4}$ are commonplace.
However, such grains are not expected to rotate suprathermally and thus
we do not discuss them there.} (Spitzer \& Tukey 1951). Therefore alignment
due to the Rowland effect is negligible for the majority of the foreseeable
cases. This a fortiori true, as  paramagnetic grains obtain much larger
magnetic moments through the Barnett, rather than through Rowland effect.
Assuming that $\omega_{T}\approx 10^{5}$~s$^{-1}$,
$B \approx 3\times 10^{-6}$~G, $\omega_{r}\approx 10^{4}\omega_{T}$,
we still fall short by approximately six orders of magnitude
to produce any measurable alignment due to this effect.

  To summarize, magnetic
moments irrespectively of their nature (the Rowland effect, the Barnett
effect, permanent magnetization) do produce alignment, but this alignment
is negligible for suprathermally rotating grains in diffuse clouds.

\section{Conclusions}

Several mechanisms of alignment of suprathermally rotating grains were
discussed and it was shown that

I. Paramagnetic alignment stays the strongest candidate to account for
 grain alignment
over vast regions of the diffuse clouds. Its modification suggested
in Purcell (1979) depends on subtle processes
over grain surfaces and this provides an opportunity to test it.

II. Alignment of grains due to the cross section effect that was introduced
in this paper can also be important for the diffuse ISM. This mechanism,
similar to the
paramagnetic one,
tends to align
 long grain axis perpendicular to magnetic field lines, provided that
the grain drift is caused by Alfv\'{e}nic perturbations.

III. Radiation can drive alignment in the vicinity of bright sources even
in the absence of supersonic drift. Such an alignment is more likely to
arise from
differential photodesorption, although in some particular cases
anisotropic radiation field can also
provide the alignment of ''helical`` grains. Therefore additional care is
required
in interpreting the corresponding polarimetric data.

IV. Magnetic moments of grains both arising from
 the Barnett effect and due to their charge  produce alignment,
but the expected degree of  alignment is
negligible.

{\bf Acknowledgement}\\ The initial impetus for this work came from
comments
by B.~Draine. I also would like to thank him for a subsequent discussion of
the results and for numerous suggestions, remarks and comments, which
considerably improved
 the paper.
  This work also owes much to my discussions with A.~Goodman,
R.~Kulsrud, P.~Myers, M.~Rees, L.~Spitzer
 and D.~Williams. The research was partially supported by Isaak Newton
Scholarship from IoA, University of Cambridge (UK), Visiting Fellowship at
CfA,  and NASA grant NAG5 2773.

\end{document}